**Possible mechanisms underlying time perception: decoupling internal and external time**

Irina Kareva and Georgy Karev

Email: ikareva@asu.edu

**Abstract**

Alignment between subjective sense of time and chronological time can become skewed as a result of pharmacological interventions, neurodegenerative diseases (such as Parkinson's disease), or even in the moments preceding brain death. Despite increased understanding of mechanisms governing time perception and the activity of the "internal clock" (such as the functionality of the dopamine system in the basal ganglia of the midbrain), there currently exist no mathematical models that allow investigation of changes in time perception through formalizing the decoupling of "internal" and "external" time. Here we propose such a model using a parametrically heterogeneous power equation, and use it to investigate the critical case of indefinite increase in internal time following cardiac arrest and preceding brain death. We identify three critical parameters that determine time to brain death and provide an analysis of the relevant quantities. We hope that our model can lay foundation for further mathematical and theoretical exploration of this complex topic.

**Introduction**

Subjective sense of time can be affected temporarily or permanently due to changes in synchronization between an "internal clock" and external "subjective" time. Arguably the most influential internal clock model, based on scalar expectancy theory (SET) was proposed by Gibbon (Gibbon 1977; Gibbon et al. 1984). According to SET, temporal processing is regulated by a pacemaker, switch and accumulator. The switch, controlled by attention, regulates the number of pacemaker pulses (clock ticks) that are collected into the accumulator. Time estimates depend on a number of "pulses" that have been accumulated between the switches: the more pulses have accumulated, the longer the perceived time interval.

This model has since been modified and expanded, and the currently accepted most plausible model of time perception, the striatal beat frequency mode (SBF), which has been proposed by Matell and Meck (2000, 2004). In the SBF, timing is based on the coincidental activation of medium spiny neurons by cortical neural oscillators in the basal ganglia in the midbrain. A broad array of studies have shown that injections of substances that act as dopamine antagonists, such as cocaine and metamphetamine, cause shorter tasks to be perceived as long, slowing down the perception of time, while drugs that activate dopamine receptors, such as haloperidol and pimozide, speed up the perception of time (Meck 1996; 2005;Coull et al. 2011). Moreover, patients with Parkinson's disease, which involves degeneration of dopaminergic substances in the basal ganglia, and specifically in substantia nigra par compacta (SNc), exhibit impaired timing perception (Malapani et al. 1998, Rammsayer and Classen 1997). Similar problems with time perception have also been observed in patients with schizophrenia (Davalos et al. 2001; Penney et al. 2005) and attention deficit hyperactivity disorder (Levy and Swanson 2001; Barkely et al. 2001). These studies suggest that there exists an "optimal" level of dopamine that aligns one's internal clock with external time.

Noticeably, temporary changes to time perception can be affected by emotions (Droit-Volet and Meck 2007; Droit-Volet 2013) and reactions to threatening situations, where involvement of the amygdala in emotional memory contributes to creation of secondary encoding of memories. This can create an erroneous a sense of events spanning a greater period of time than has actually passed and appears to be a function of recollection, not perception of time (Stetson et al. 2007).

Another particularly interesting case of alterations in time perception could be occurring during near-death experiences (French 2005). Anecdotal evidence suggests that time intervals during a near-death experience are perceived to be much longer than they actually are. Moreover, in 2013, Borjigin et al. reported a high frequency neurophysiological activity in the near-death state of rats undergoing experimental cardiac arrest. The levels of neurophysiological activity in the near-death state exceeded those found during conscious waking state, suggesting that the brain in fact might be highly active in the near-death state.

These observations, coupled with the current state of knowledge about changes time perception, allow formulation of the following hypothesis: During a near-death experience, lack of oxygenation eventually causes death of all brain cells, including the dopaminergic substances in the basal ganglia, causing an experience of internal time becoming increasingly longer compared to external time. This could be a possible mechanism to account for the perception of time that allows "life to flash before one's eyes" in the moments preceding death.

In what follows, we propose a mathematical model that allows decoupling of internal and external time. We propose an interpretation, which allows simulating internal time tending to infinity in finite time interval, recapitulating qualitatively the experience of time perception in a dying brain. While we understand that this is solely a theoretical conceptual model, we hope that it might provide insights into mechanisms that could underlie time perception.

**Methods**

We propose a conceptual mathematical model that allows investigation of a population of neurons responsible for changes in time perception. We chose a power equation for population decrease because of the phenomenological properties that these types of models allow, which we deem to be appropriate for this particular investigation.

First, we will introduce the model in its general form, and discuss its properties. Then, we will focus on a subclass of power models and present an argument that allows to mathematically decouple what can be interpreted as "internal" and "external" time. Finally, we will demonstrate, under which conditions internal time can tend to infinity on a finite interval, replicating qualitatively a possible experience of a dying brain.

**Mathematical model**

Consider the following power model of population extinction, shown in power Equation (1), where $N$ is the population size of neural cells, and $k$ and $p$ are positive constants:

$$\frac{dN}{dt} = -kN^p \qquad (1)$$

We will provide a possible interpretation for their meaning later in this section.

The power models of population growth, $\frac{dN}{dt} = kN^p$ were introduced by E.Szathmary and M.Smith (1997). Three cases are distinguished: the *exponential* with $p=1$; the *super-exponential* with $p>1$ and the *sub-exponential* with $p<1$.

Some well-established examples of non-exponential growth apply to some molecular replicator systems (*p*=1/2, von Kiedrowski 1993) and global demography (*p*=2, van Forster 1960, Kapitza 1996, Karev 2005). It is not always clear why non-exponential growth is observed in reality. Recently it was proven in (Karev 2014; Karev and Koonin 2013) that non-exponential power models of population growth can be understood within the frameworks of inhomogeneous population models. Similar results are valid for Equation (1). We can show that any power model of extinction can be presented in a common "canonical" form as the equation for total population size of inhomogeneous "frequency-dependent model" (hereafter referred to as F-model) with a special initial distribution of the Malthusian death rate. F-models of extinction possess a number of interesting properties, which will be discussed further.

**Non-exponential power models. General results.**

Let us assume that each individual in the population (e.g., a neural cell) is characterized by a value of a parameter of its own death rate. We define $l(t,a)$ to be a clone, which here stands for a set of all individuals in the population having the death rate (Malthusian parameter) equal to *a*.

Now, assume that the death rates take the values from some set $A$. Then by definition the total population size

$$N(t) = \int_A l(t,a) \, da \qquad (2)$$

and $P(t,a) = \dfrac{l(t,a)}{N(t)}$ is the distribution of the Malthusian parameter $a$ at $t$ moment.

The following assertion is valid:

*Any power equation* $\dfrac{dN}{dt} = -kN^p$, $p > 0$ *describes the dynamics of total size of inhomogeneous frequency-dependent extinction model*

$$\frac{dl(t,a)}{dt} = -kaP(t,a) \qquad (3)$$

*where the initial distribution of the parameter $a$, $P(0,a)$, is the Gamma-distribution with the mean $N(0)^p$ and variance $pN(0)^{2p}$.*

A complete formulation is given in Appendix.

Equation (3) gives a canonical form of the power model (1) for any $p$.

We can also show (see Appendix) that *the solution to the power Equation (1) exists for all $t > 0$ if $p \geq 1$ and only up to the moment of population extinction $T = N(0)^{1-p} / (k(1-p))$ when $p < 1$*. It means that the life time of populations described by exponential or super-exponential equation is indefinite. In contrast, a sub-exponential population goes to extinction not asymptotically, as super-exponential power models, but at certain finite time moment, i.e. when $N(T) = 0$.

For this reason, the sub-exponential power models are of primary interest within the context of the proposed investigation.

**Internal and external time**

In what follows we use the sub-exponential version of power Equation (1) with $p < 1$ in order to construct an inhomogeneous model of population extinction such that the total population size solves Equation (1). We choose the sub-exponential growth model because this type of model will have the properties that are necessary for our question.

Let us make a formal change of time $t \to q$ in Equation (3) by the equation

$$dq = \frac{k}{N(t)} dt. \tag{5}$$

Then Equation (3), $\frac{dl(t,a)}{dt} = -ka \frac{l(t,a)}{N(t)}$, with respect to this new independent variable $q$ reads

$$\frac{dl(q,a)}{dq} = -al(q,a). \tag{6}$$

Equation (6) becomes a standard Malthusian equation of population extinction.

Within this framework, *the variable $q$ can be interpreted as the "internal time" of this inhomogeneous population*. Such interpretation becomes possible because *each clone (subpopulation) $l(q,a)$ with respect to this time scale evolves as if it does not depend on other clones and on the population as a whole*. Notice also that $q(t)$ is the only time scale for the Equation (3), which possesses by this property.

Remark, that any inhomogeneous F-model (3) with respect to the internal time defined by Equation (5) becomes identical to the inhomogeneous Malthusian model with respect to the "common", or "external" time, which describes free growth of clones.

The internal time $q(t)$ as function of "real time" $t$ is given in Appendix, Theorem 1,s.3):

$$q(t) = \frac{(1 - kN(0)^{p-1}(1-p)t)^{\frac{p}{p-1}} - 1}{pN(0)^p}. \tag{7}$$

As one can easily see, it follows from this formula that *if $p \geq 1$, then $q(t)$ is finite for all $t$, and $q(t) \to \infty$ when $t \to \infty$*. In contrast, if $0 < p < 1$, then $q(t) \to \infty$ as $t \to T < \infty$, where $T$ is defined by

$$T = \frac{N(0)^{1-p}}{k(1-p)} \tag{8}$$

We may interpret $T$ as the extinction moment because at the moment $T$ the population completely goes to extinction, i.e. $N(T) = 0$.

Therefore, for $p < 1$, the value of $T$, the time point at which internal time $q(t)$ tends to infinity, depends largely on the initial population size $N(0)$, while for $p \geq 1$, this effect is no longer present. Within the context of our model, this suggests that the time to $T$ depends on the initial number of neurons, or connections between them, both of which are logically consistent.

Noticeably, *a finite* "real time" interval $[0, T)$ for sub-exponential inhomogeneous model in Equation (3) corresponds to *infinite* duration of "internal" time $q(t)$.

A typical plot of $q(t)$ for $p < 1$ is shown in Figure 1.

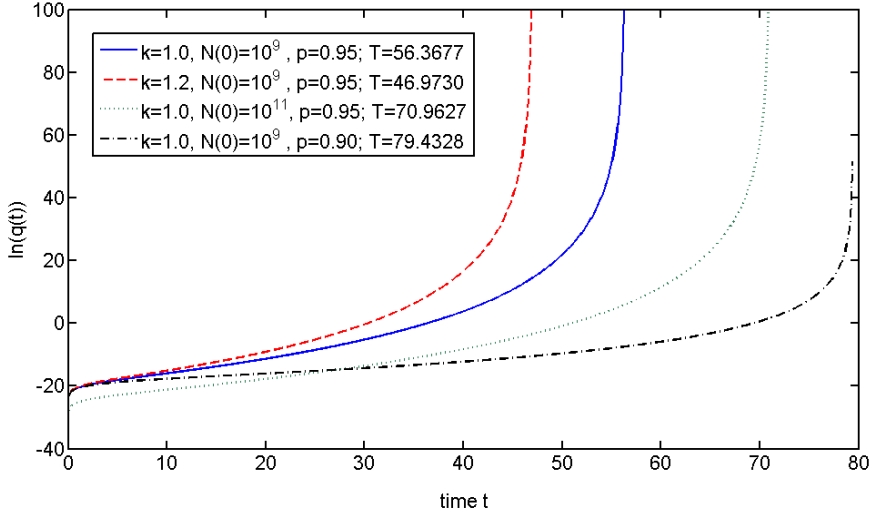

**Figure 1.** Plot of internal time $q(t)$, defined in Equation (7), for $p < 1$, in logarithmic scale, against real time.

As one can see, larger values of $k$ decrease extinction time $T$, defined in Equation (8). Increasing the initial value of $N(0)$ predictably delays extinction time $T$. Noticeably, dependence of $T$ on the value of $p$ is non-monotonic, as can be seen in Figure 2.

*Units and possible interpretations for variables and parameters*

Here, we propose that $q(t)$, defined in Equation (7), is a possible representation of internal time. From that equation it follows that $T$ is the time moment of population extinction, which occurs as $q(t) \to \infty$ and $t \to T < \infty$.

There are three parameters that determine the dynamics of the internal time $q(t)$: $k$, $N(0)$ and $p$. The units for these three quantities are $k: \dfrac{(cells)^{(1-p)}}{time}$, $p$ is unitless, and $N$ can be interpreted as either the number of cells, or possibly the number of connections between cells. The latter interpretation might be preferable, since demise of connections between cells in a damaged or dying brain can lead to different parts of the brain living and dying as if separate from each other, which underlies the assertion that $q(t)$ can in fact be representing internal time. An average human brain is estimated to range between 85 and 120 billion (Herculano-Houzel 2009) neurons, so we can estimate $N(0)$ to be roughly $10^{11}$. The number of synapses (based on 1000 per neuron estimate) is about $10^{14}$, or 100 trillion.

Furthermore, from Equation (7), the expression $kN(0)^{(p-1)}$ can be seen as a scaling quantity for "external" time, and has the dimension $\frac{1}{time}$.

For the power $p$, we do not yet have a well-established biological interpretation but we can identify certain properties. Specifically, if $p > 1$, the population defined in power Equation (1) lives forever, but for $p < 1$ it will go extinct in finite time, defined by $T$. Hence, the parameter $p$ can be estimated based on the duration of the process of population extinction. Equation (9) below gives a simple estimation of $p$ with respect to the initial population size $N(0)$.

The inhomogeneous model in Equation (3) noticeably does not depend on $p$. According to Equation (3), the rate of extinction of cells, or synaptic connections between them, is proportional to their frequency. However, the distribution of parameter *a*, which defines population clones, does depend on $p$. Specifically, the expected value and the variance of parameter $a$, which define completely the initial Gamma-distribution, both depend on the value of $p$. Therefore theoretically, the mean value of the parameter *a* can be used for estimation of the power *p*, as given by $E^t[a] = \rho\beta(t) = N(t)^p$ (see Equation (A.19) and corresponding derivation in Appendix).

**Analysis**

From Equation (8), one can easily show that $T$ as a function of $p$ has a single minimum (see Figure 2), and the minimal value of $T$ at given $N(0)$ is attained when

$$p = p_{opt} = 1 - \frac{1}{lnN(0)}. \tag{9}$$

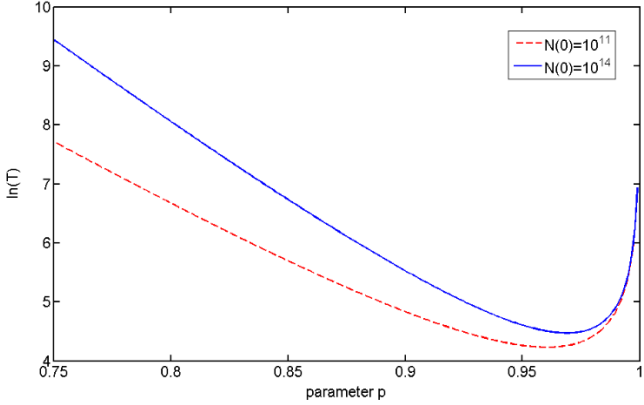

**Figure 2.** The plot of extinction moment $T$, defined in Equation (8), plotted in logarithmic scale against the power $p$ for $N(0) = 10^{11}$ (red) and $N(0) = 10^{14}$.

As one can see, there exists a single value of $p = p_{opt}$, which minimizes $T$ (for instance, $p_{opt} = 0.96$ for $N(0) = 10^{11}$ and $p_{opt} = 0.97$ for $N(0) = 10^{14}$).

Furthermore, by plotting $p_{opt}$ vs $N(0)$ according to Equation (9), we can see that for a very large magnitude of values of $N(0)$, $10^{10} < N(0) < 10^{15}$, the values of $p_{opt}$, which provide minimum of $T$ at given initial population size, are $0.957 < p_{opt} < 0.971$, demonstrating that the "optimal" value of $p$ belongs to a very narrow range.

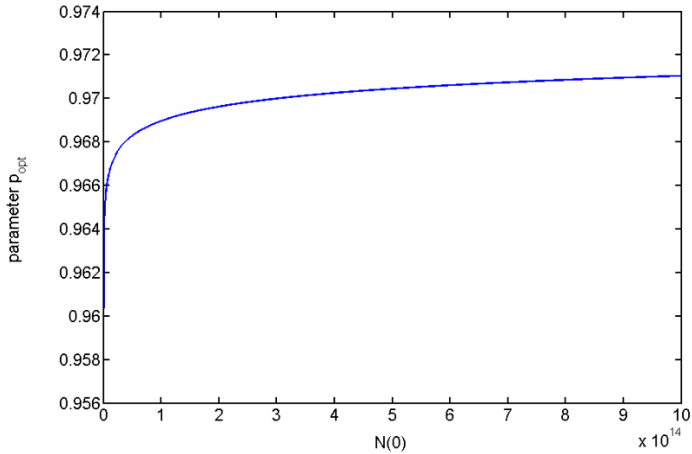

**Figure 3.** The plot of the optimal value $p_{opt}$ given by Equation (9), against $N(0)$, plotted for $10^{10} < N(0) < 10^{15}$.

For $p_{opt}$ defined by Equation (7), the minimal value of $T$ is equal to

$$T_{min} = \frac{lnN(0) N(0)^{\frac{1}{lnN(0)}}}{k} = \frac{e}{k} lnN(0) \qquad (10)$$

since $N^{\frac{1}{lnN}} \equiv e$. A plot of $T_{min}$ vs $N(0)$ is given in Figure 4.

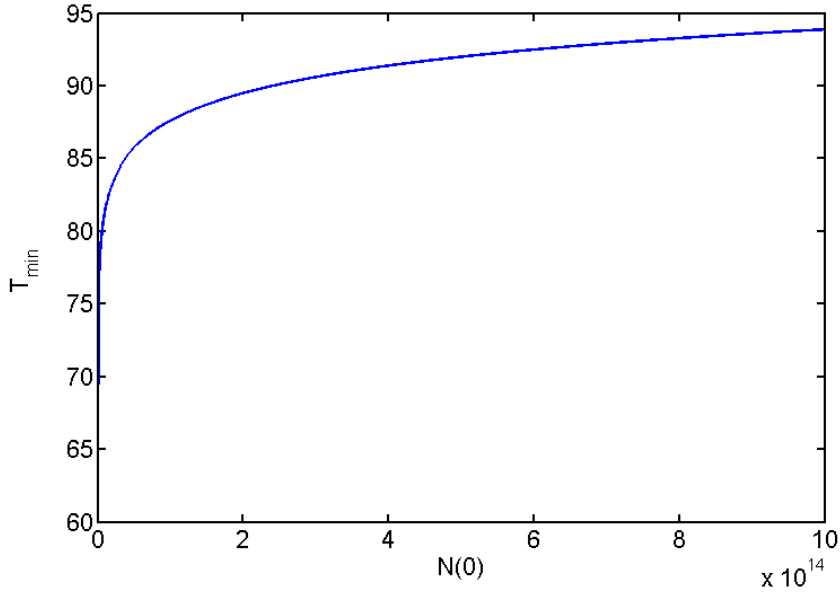

**Figure 4**. Plot of $T_{min}$, defined in Equation (10), against $N(0)$, $10^{11} < N(0) < 10^{14}$; $k = 1$.

*Interpretation in the context of the human brain*

The duration of death of the brain core in the state of clinical death is between 5 and 20 minutes (Safar 1988).

For instance, if $T = 300\,sec$, $N(0) = 10^{11}$, then according to Equation (10), $k = 0.23$; if $N(0) = 10^{14}$, then $k = 0.29$.

Then, within the framework of our model, if the process of death of brain core can be described by power equation (1), $\frac{dN}{dt} = -kN^p$, with $10^{11} < N(0) < 10^{14}, 0.96 < p < 0.97$, and $0.23 < k <$ 0.29, then during $T = 5$ min of "external", or chronological time, the "internal" time of the underlying *F*-model can become indefinitely large, and potentially infinite.

**Discussion**

Perception of time and alignment of one's "internal clock" with "external" chronological time appears to be primarily affected by the activity of the dopamine in basal ganglia in the midbrain (Matell and Meck 2000, 2004). Specifically, it has been shown that substances that act as dopamine antagonists cause shorter time periods to appear long; this effect is reversed by substances that activate dopamine receptors (Meck 1996; 2005; Coull et al. 2011). Moreover, patients with diseases that involve degeneration of dopaminergic substances in the basal ganglia, such as Parkinson's disease, exhibit similar impairments in time perception (Malapani et al. 1998, Rammsayer and Classen 1997). We hypothesize that when a brain is dying, such as after a cardiac arrest, the eventual cessation of activity of all neural cells, including dopamine system in the basal ganglia, can cause a perception of time to increase dramatically compared to "external" chronological time.

Here we proposed a mathematical model that allowed us to replicate this effect through decoupling "internal" and "external" time using a parametrically heterogeneous sub-exponential power equation, where cells, or possibly synapses, $N(t)$ in the population die at different rates. Specifically, we divided the population of neural cells into clones, and death rate of each clone is proportional to its frequency in the population.

Through a series of transformations, we derived Equation (7) for $q(t)$, a variable that can be interpreted as describing "internal time", because each clone, or subpopulation, with respect to this time scale evolves as if it does not depend on other clones or on the population as a whole. Within a context of a dying brain, this would describe a situation when loss of connections between cells in the dying brain would indeed lead to subpopulations of cells dying independently of each other, which is logically consistent. Furthermore, from this equation we were able to also specify time of death $T$, where population of cells goes to extinction , i.e., when $N(T) = 0$. Conversely, knowing the specific time of death $T$, we can estimate other model parameters.

There were three parameters governing the dynamics of the internal time $q(t)$, namely $k$, $p$ and $N(0)$. $N(0)$ represents the initial number of cells, or alternatively, connections between cells; $k$ determines the death rate of a population as a whole and is involved in a scaling quantity that connects internal time to external time. Parameter $p$, the power in the non-exponential Equation (1) of population growth, crucially defines the time of population extinction $T$. The power $p$ has to be $p < 1$ because in other case $T$ is indefinite, rendering the population $N(t)$ immortal. Increase in the value of parameter $k$ decreases time to extinction $T$, defined in Equation (8). Increase in the initial value of $N(0)$ predictably delays extinction time $T$ (Figure 1). Dependence of $T$ on the value of $p$ is non-monotonic, and there exists a $p_{opt}$ for each $N(0)$ (Figure 2), which minimizes time $T$. Moreover, we proved that $0.957 < p_{opt} < 0.971$ for $10^{10} < N(0) < 10^{15}$, which is the estimated number of neural cells in the brain or even the number of synapses (Herculano-Houzel 2009). The result in visualized in Figure 3. Finally, we also derived an equation for $T_{min}$, the minimal value of $T$ provided by $p_{opt}$ (Figure 4).

One can apply this model to explore hypotheses about what may be happening with time perception in the moments preceding death. Given that the malfunction of the dopamine system in the basal ganglia is responsible for alterations in time perception (Mattell and Meck 2000; 2004), then within the frameworks of this model, death of cells, or synapses $N(t)$ will result in internal time $q(t) \to \infty$ in a finite period of time, denoted by $T$. The details of what happens with time perception prior to brain death are not yet understood, including whether internal time indeed increases in the moments preceding death, allowing one to experience more than would normally be possible (Bierce 2008). Here we propose one possible hypothesis which is subject to future investigation.

It is not known what mechanisms could be governing perception of time in the human brain. Here we proposed a possible mathematical formalization, which provides predictions that are logically consistent. We have identified a small number of key parameters that could be involved in the decoupling of "internal" and "external" time in the human brain, and proposed an explanation for mechanisms underlying time perception. It is our hope that this model can lay a foundation for further mathematical and theoretical exploration of this complex topic, which eventually might yield results to deepen the understanding of diseases, such as Parkinson's, schizophrenia and ADHD, where accurate time perception is compromised.

**Acknowledgements**

This research was conducted in 2015. It was supported by the Intramural Research Program of the NIH, NCBI. The views presented here are the authors' personal views.This research was conducted in 2015. It was supported by the Intramural Research Program of the NIH, NCBI. The views presented here are the authors' personal views.

**Mathematical Appendix**

**Theorem**.

1) Any power equation

$$\frac{dN}{dt} = -kN^p, \quad p > 0 \tag{A.1}$$

describes the total size of inhomogeneous frequency-dependent model

$$\frac{dl(t,a)}{dt} = -\frac{kal(t,a)}{N(t)} = -kaP(t,a), \tag{A.2}$$

where the initial distribution of the parameter $a$, $P(0,a)$ is the Gamma-distribution

$$P(a) = \frac{a^{\rho-1}\exp(-\frac{a}{\beta})}{\beta^\rho \Gamma(\rho)}, \quad a > 0, \tag{A.3}$$

with $\rho = \frac{1}{p}$ and $\beta = \beta(0) \equiv pN(0)^p$;

2) the solution to the power equation (A.1) exists for all $t > 0$ at $p \geq 1$ and only up to the moment of population extinction

$$T = \frac{1}{(kN(0)^{p-1}(1-p))} \quad at \ p < 1, \tag{A.4}$$

at this moment $N(T) = 0$;

3) the solution to the inhomogeneous F-model (A.2) is

$$l(t,a) = l(0,a)e^{-aq(t)},$$

where

$$q(t) = \frac{(1-kN(0)^{-1+p}(1-p)t)^{\frac{p}{1-p}} - 1}{pN(0)^p} \quad \text{for } p \neq 1 \tag{A.5}$$

and $q(t) = \dfrac{e^{kt} - 1}{N(0)}$ for $p = 1$;

4) *the total population size at the moment t*

$$N(t) = N(0)\left(1 - kN(0)^{-1+p}(1-p)t\right)^{\frac{1}{1-p}} \tag{A.6}$$

for $p \neq 1$ and $N(t) = N(0)e^{-kt}$ for $p = 1$.

5) *the current distribution of the parameter a at moment t is Gamma-distribution (A.3) with*

$\rho = \dfrac{1}{p}$ and $\beta(t) = pN(t)^p$; *its mean is* $E^t[a] = \rho\beta(t) = N(t)^p$ *and the variance*

$Var^t[a] = \rho\beta(t)^2 = pN(t)^{2p}$.

**Proof.**

Let us construct the inhomogeneous model (A.2) such that its total population size solves equation (A.1).

Let $\dfrac{dl(t,a)}{dt} = -\dfrac{kal(t,a)}{N(t)} = -kaP(t,a)$. Then

$$\dfrac{dN}{dt} = \int_A \dfrac{dl(t,a)}{dt} da = -k\int_A aP(t,a)da = -kE^t[a] \tag{A.7}$$

In order to obtain Equation (A1) we need to have $E^t[a] = N^p$.

To this end, let us define formally the auxiliary variable by the equation

$$\dfrac{dq}{dt} = kN^{-1}, \quad q(0) = 0. \tag{A.8}$$

Then $l(t,a) = l(0,a)e^{-aq(t)}$.

Define the Laplace transform of the initial distribution of the parameter $a$

$$L_0(\delta) = \int_A e^{-\delta a} P(0,a) da \qquad (A.9)$$

Then

$$N(t) = \int_A l(t,a) da = N(0) L_0(q(t)), \qquad (A.10)$$

$$P(t,a) = \frac{e^{-aq(t)}}{L_0(q(t))} P(0,a), \qquad (A.11)$$

$$E^t[a] = \int_A a P(t,a) da = -\frac{d}{dq} \ln L_0(q(t)) \qquad (A.12)$$

Next, if we wish to have $E^t[a] = N(t)^p = N(0)^p L_0(q(t))^p$, then the following equations should hold:

$$E^t[a] = -\frac{d}{dq} \ln L_0(q) = -\frac{1}{L_0(q)} \frac{dL_0(q)}{dq} = N(0)^p L_0(q)^p. \qquad (A.13)$$

Hence,

$$\frac{dL_0(q)}{dq} = -N(0)^p L_0(q)^{p+1}. \qquad (A.14)$$

Solving this equation we obtain

$$L_0(q) = (1 + pN(0)^p q)^{-\frac{1}{p}}. \qquad (A.15)$$

Recall that $L_0$ should be the Laplace transform of an initial distribution of the parameter $a$. It is well known that the function $L_0(\delta) = (1 + \beta \delta)^{-\rho}$ for $\beta > 0$ is the Laplace transform of the Gamma-distribution (A.3) The mean and variance of this distribution are $\rho \beta$ and $\rho \beta^2$.

Hence, the initial distribution of the parameter $a$ should be the Gamma-distribution (A.16) with $\rho = \dfrac{1}{p}$ and $\beta = pN(0)^p$.

Now we can compute the auxiliary variable $q(t)$, as well as all statistical characteristics of interest. Since $N(t) = N(0)L_0(q(t))$, then

$$\frac{dq}{dt} = kN^{-1} = k(N(0)L_0(q))^{-1} = k(N(0))^{-1}(1 + pN(0)^p q)^{\frac{1}{p}}. \tag{A.16}$$

The solution to this equation under initial condition $q(0) = 0$ is

$$q(t) = \frac{(1 - kN(0)^{p-1}(1-p)t)^{\frac{p}{p-1}} - 1}{pN(0)^p} \qquad \text{for } p \neq 1 \tag{A.17}$$

and $q(t) = \dfrac{e^{kt} - 1}{N(0)}$ for $p = 1$.

The Laplace transform of the current distribution $P(t, a)$ given by (A.11)

$$L_t(\delta) = E^t\left[e^{-\delta a}\right] = \frac{L_0(\delta + q(t))}{L_0(q(t))} = (1 + \frac{pN(0)^p \delta}{1 + pN(0)^p q(t)})^{-\frac{1}{p}} = (1 + pN(t)^p \delta)^{-1/p}; \tag{A.18}$$

It is the Laplace transform of the Gamma-distribution (A.16) with $\rho = 1/p$ and $\beta(t) = pN(t)^p$.

Its mean value is given by

$$E^t[a] = \rho\beta(t) = N(t)^p \tag{A.19}$$

as desired.

The population size is given by

$$N(t) = N(0)L_0(q(t)) = N(0)(1 - kN(0)^{-1+p}(1-p)t)^{\frac{1}{1-p}} \qquad \text{for } p \neq 1 \tag{A.20}$$

and $N(t) = N(0)e^{-kt}$ for $p = 1$.

Hence, if $p < 1$, then $N(t) \to 0$ as $t \to T$, where

$$T = \frac{1}{kN(0)^{-1+p}(1-p)}. \tag{A.21}$$

All assertions of the Theorem are proven.

**References**

Barkley, R. A., Murphy, K. R., & Bush, T. (2001). Time perception and reproduction in young adults with attention deficit hyperactivity disorder. *Neuropsychology*, *15*(3), 351.

Bierce, A. (2008). *An occurrence at Owl Creek Bridge and other stories*. Courier Corporation.

Borjigin, J., Lee, U., Liu, T., Pal, D., Huff, S., Klarr, D., ... & Mashour, G. A. (2013). Surge of neurophysiological coherence and connectivity in the dying brain. *Proceedings of the National Academy of Sciences*, *110*(35), 14432-14437.

Coull, J. T., Cheng, R. K., & Meck, W. H. (2011). Neuroanatomical and neurochemical substrates of timing. *Neuropsychopharmacology*, *36*(1), 3-25.

Davalos, D. B., Kisley, M. A., & Ross, R. G. (2002). Deficits in auditory and visual temporal perception in schizophrenia. *Cognitive Neuropsychiatry*, *7*(4), 273-282.

Droit-Volet, S. (2013). Time perception, emotions and mood disorders. *Journal of Physiology-Paris*, *107*(4), 255-264.

Droit-Volet, S., & Meck, W. H. (2007). How emotions colour our perception of time. *Trends in cognitive sciences*, *11*(12), 504-513.

French, C. C. (2005). Near-death experiences in cardiac arrest survivors. *Progress in Brain Research*, *150*, 351-367.

Forster, von H. et al. (1960) Doomsday: Friday, 13 November, A.D. 2026. *Science* 132: 1291-99.

Gibbon, J., Church, R. M., & Meck, W. H. (1984). Scalar timing in memory. *Annals of the New York Academy of sciences*, *423*(1), 52-77.

Gibbon, J. (1977). Scalar expectancy theory and Weber's law in animal timing. *Psychological review*, *84*(3), 279.

Herculano-Houzel, S. (2009). The human brain in numbers: a linearly scaled-up primate brain. *Frontiers in human neuroscience*, *3*.

Kapitza S. (1996) The phenomenological theory of world population growth. *Physics-Uspekhi* 39, 1: 39-57.

Karev, G. P. (2005). Dynamics of inhomogeneous populations and global demography models. *Journal of Biological Systems*, *13*(01), 83-104.

Karev, G. P., & Koonin, E. V. (2013). Parabolic replicator dynamics and the principle of minimum Tsallis information gain. *Biol. Direct*, *8*, 19-19.

Karev, G. P. (2014). Non-linearity and heterogeneity in modeling of population dynamics. *Math Biosciences, 258,* 85-92.

von Kiedrowski, G. (1993). Minimal replicator theory I: Parabolic versus exponential growth. In *Bioorganic chemistry frontiers* (pp. 113-146). Springer Berlin Heidelberg.

Levy, F., & Swanson, J. M. (2001). Timing, space and ADHD: the dopamine theory revisited. *Australian and New Zealand Journal of Psychiatry*, *35*(4), 504-511.

Malapani, C., Rakitin, B., Levy, R. S., Meck, W. H., Deweer, B., Dubois, B., & Gibbon, J. (1998). Coupled temporal memories in Parkinson's disease: a dopamine-related dysfunction. *Cognitive Neuroscience, Journal of*, *10*(3), 316-331.

Matell, M. S., & Meck, W. H. (2000). Neuropsychological mechanisms of interval timing behavior. *Bioessays*, *22*(1), 94-103.

Matell, M. S., & Meck, W. H. (2004). Cortico-striatal circuits and interval timing: coincidence detection of oscillatory processes. *Cognitive brain research*, *21*(2), 139-170.

Meck, W. H. (1996). Neuropharmacology of timing and time perception.*Cognitive brain research*, *3*(3), 227-242.

Meck, W. H. (2005). Neuropsychology of timing and time perception. *Brain and cognition*, *58*(1), 1-8.

Penney, T. B., Meck, W. H., Roberts, S. A., Gibbon, J., & Erlenmeyer-Kimling, L. (2005). Interval-timing deficits in individuals at high risk for schizophrenia.*Brain and cognition*, *58*(1), 109-118.

Rammsayer, T., & Classen, W. (1997). Impaired temporal discrimination in Parkinson's disease: temporal processing of brief durations as an indicator of degeneration of dopaminergic neurons in the basal ganglia. *International Journal of Neuroscience*, *91*(1-2), 45-55.

Safar, P. (1988). Resuscitation from clinical death: pathophysiologic limits and therapeutic potentials. *Critical care medicine*, *16*(10), 923-941.

Stetson, C., Fiesta, M. P., & Eagleman, D. M. (2007). Does time really slow down during a frightening event. *PLoS One*, *2*(12), e1295.

Szathmáry, E., & Smith, J. M. (1997). From replicators to reproducers: the first major transitions leading to life. *Journal of theoretical biology*, *187*(4), 555-571.